\documentclass[conference, onecolumn]{IEEEtran}
\usepackage{mathtools}
\usepackage{amsmath}
\usepackage{blindtext}
\usepackage{bbm}
\usepackage{comment}
\usepackage{graphicx}
\usepackage[ruled,vlined]{algorithm2e}

\usepackage{amsmath,xcolor,amssymb,multirow,cite,epsfig,graphicx,graphics,citesort}
\usepackage{subcaption}
\renewcommand{\baselinestretch}{1.2}

\begin{document}
\title{Wildfire Monitoring in Remote Areas using Autonomous Unmanned Aerial Vehicles}

\author{
	\IEEEauthorblockN{Fatemeh Afghah\IEEEauthorrefmark{1},  Abolfazl Razi\IEEEauthorrefmark{1}, 
Jacob Chakareski\thanks{Distribution A: Approved for Public Release, distribution unlimited.  Case Number 88ABW-2019-1454 on 03 Apr. 2019. The work of F. Afghah was supported by AFRL. The work of A. Razi was supported by NSF Award CNS-1755984. The work of J. Chakareski was supported in part by NSF Awards CCF-1528030, ECCS-1711592, CNS-1836909, and CNS-1821875.}\IEEEauthorrefmark{2}, Jonathan Ashdown\IEEEauthorrefmark{3}}

	\IEEEauthorblockA{\IEEEauthorrefmark{1}School of Informatics, Computing and Cyber Systems, Northern Arizona University, Flagstaff, AZ, USA}
         
	\IEEEauthorblockA{\IEEEauthorrefmark{2}Electrical and Computer Engineering, University of Alabama,
Tuscaloosa, AL, USA}
	\\

\IEEEauthorblockA{\IEEEauthorrefmark{3} Air Force Research Laboratory, Rome, NY, USA} 
}


\maketitle
\thispagestyle{plain}
\pagestyle{plain}
\begin{abstract}
In this paper, we propose a drone-based wildfire monitoring system for remote and hard-to-reach areas. This system utilizes autonomous unmanned aerial vehicles (UAVs) with the main advantage of providing on-demand monitoring service faster than the current approaches of using satellite images, manned aircraft and remotely controlled drones. Furthermore, using autonomous drones facilitates minimizing human intervention in risky wildfire zones. 
In particular, to develop a fully autonomous system, we propose a distributed leader-follower coalition formation model to cluster a set of drones into multiple coalitions that collectively cover the designated monitoring field. The coalition leader is a drone 
that employs observer drones potentially with different sensing and imaging 
capabilities to hover in circular paths and collect imagery information from the impacted areas. The objectives of the proposed system include: 
i) to cover the entire fire zone with a minimum number of drones, and ii) to minimize the energy consumption and latency of the available drones to fly to the fire zone. 
Simulation results confirm that the performance of the proposed system- without the need for inter-coalition communications- approaches that of a centrally-optimized system.
\footnote{Distribution A: Approved for Public Release, distribution unlimited.  Case Number 88ABW-2019-0315. }.

\end{abstract}
\IEEEpeerreviewmaketitle

\section{Introduction}
Wildfires are one of the costliest and deadliest natural disasters across the world, especially in the US West region. The immediate impacts include damage to millions of hectares of forest resources, evacuation of thousands of people, burning of homes and devastation of infrastructure, and most importantly, threatening the lives of people \cite{NIST_fire}. Moreover, wildland fires may also disrupt forestry operations, which produce wood fiber and biomass fuels, and other forms of agriculture, by spreading into farms and damaging ecosystems with negative consequences on water quality and other ecosystem services \cite{fire_eco}.
National institute of standards and technology (NIST) estimates that the annual cost of wildfire management ranges from \$7.6 billion to \$62.8 billion, while the annual losses caused by wildfires are much higher, ranging from \$63.5 billion to \$285.0 billion \cite{NIST_fire}. In the US, wildfire suppression costs exceeded \$2 billion in 2017, breaking the all-time record \cite{USDA_17}. 
The recent fire in California during November 2018 that has been the largest in the history of this state has killed 88 people, and burned over 600,000 acres. The rapidly increasing risk of fire, due to recent widespread extreme drought conditions and climate change, calls for new national strategies to prevent and manage wildfires, at a reasonably low cost. 

Several technologies have been used for fire detection and monitoring including ground sensors, remotely piloted vehicles (RPV), or satellite imaging \cite{Allison}. However, these methods are not yet able to offer a fast and reliable solution for wildfire detection and monitoring \cite{Enrico,Enrico2}. 
Some drawbacks of the current technologies include: i) delayed fire detection due to missing small fires at early stages, ii) relatively long time lag for satellites to overpass the field, and iii) infeasibility of deploying sensors with limited sensing distance ranges (e.g., chemical-based smoke detectors). For instance, smoke detectors are efficient in detecting fires at early stages, but they suffer from short distance ranges. Imagery systems in the visible light spectrum can be used for remote sensing, but they lack accuracy at nighttime, and cloudy and foggy weather conditions. 
More importantly, current forest fire suppression and management techniques involve ground-based personnel and aircraft pilots that put them at risk and is costly to operate. 

UAVs are presently banned from operations in wildfire zones due to potential collisions with aircraft flying at low altitudes \cite{NIFC}. However, drone-based wildfire monitoring can provide a low cost, and rapid imaging solution specially in low populated areas. UAVs have been recently tested in several fire monitoring missions \cite{Allison,drone_fire3,fire_survey}. 
However, in these examples, a single remotely-controlled UAV is utilized in a line of sight range of the human controller, which still highly involves human intervention and may seriously endanger the lives of controllers. Other disadvantages of using single monitoring drone includes (i) low spatial and temporal resolution, and (ii) limited flight time of a single UAV (often less that 45 minutes).

A European project named \textit{real-time coordination and control of multiple heterogeneous UAVs} (COMETS) has focused on utilizing a fleet of heterogeneous drones for forest surveillance, and forest fire detection and observation with the goal of reducing the operation cost using less costly small drones and enhancing the imaging resolution \cite{Martínez-de-Dios2007,Cruz,Merino}. However, in this project, the drones are controlled by human controllers that endangers the life of controllers and limits the use of system in remote and hard-to-reach regions.  

In this paper, we propose a practical framework for fire monitoring using a set of heterogeneous autonomous drones in inaccessible regions to enable rapid mission response when human controllers are not available to initiate and guide the mission.
This method is based on a decentralized leader-follower coalition formation to provide a full coverage of the fire zone with a limited number of available UAVs.

\section{Review of Related Work}
UAV-based remote sensing is an emerging technology that has been utilized in a wide range of civilian and military applications including environmental monitoring, and precision agriculture \cite{BEKMEZCI20131254,Adams_survey,Peng_UAV,Alireza_CCNC19,Mehrdad_SECON18}. Similarly, UAV-enabled aerial small cells have been explored to extend the capacity, coverage, and energy efficiency of 5G heterogeneous cellular networks featuring millimeter wave multi-band and multi-tier network architectures \cite{Jacob_Green2019,Jacob_ICC18,Korenda_CISS,Razi_packet_TWC16}. Finally, \cite{Chakareski:17,Chakareski:17a} investigate UAV-IoT data acquisition, networking, and path planning towards enabling next generation applications such as networked virtual and augmented reality, where edge computing is integrated \cite{Chakareski:17b}, and effective allocation of the wireless bandwidth across the multiple sensing locations needs to be carried out \cite{ChakareskiVS:12,Chakareski:15}.

In addition to using drones with onboard sensing systems, in a different class of UAV-based sensing systems, the drones operate as a bridge between a terrestrial wireless sensor network and a cloud based processing unit, such that a single or multiple UAVs fly in pre-programmed paths and collect information from a grid of stationary sensors \cite{Razi_Asilomar17}. Although this method works efficiently for offline data aggregation and processing, it is not well suited for agile reaction to life-threatening events such as wildfire, and flood. In such natural disasters with highly dynamic operation fields, a rapid response is required. Also using drones with direct communications to ground stations, specially when the disaster occurs in far-to-reach areas, consumes a considerable portion of the UAVs' energy. In these scenarios, using autonomous UAVs with the capability of independent path planning or task completion provides a more feasible and efficient solution.     
To overcome the limited capabilities of a single UAV including limited payload, short flight time and limited communications range, a network of multiple UAVs can be utilized to cooperatively perform compound tasks and cover wide and highly dynamic operation fields. Several recent researches have been focused on the problem of centralized or decentralized task allocation in multi-agent systems 
that commonly rely on full knowledge of the agents' resources and positions, and hence require reliable communication systems among the agents to share such information \cite{Afghah_INFOCOM18,Schneider,Korsah,Behzad_RL,Behzad_Qlearning,shamsoshoara2015enhanced,shamsoshoara2019overview,mousavi2017traffic,mousavi2016learning,GHAZANFARI201661,shamsoshoara2019ring,mousavi2016deep}
These methods cannot offer an urgent response in dynamic situations, where the tasks take place at unpredictable locations and times, and a full prior knowledge of the operation field is not available. Therefore, they cannot offer a practical and real-time task allocation mechanism for UAV-based monitoring systems. A clear example of such disasters is a forest widefire where the challenges include: i) the lack of a complete  prior knowledge of the remote operation field, ii) high dynamicity and the large extent of the fields (i.e. fire spread due to wind and weather conditions), 
iii) the vulnerability of drones to malfunction and failure 
and iv) the limited sensing, computation and communication capability of a single UAV. 
These challenges urge to develop an efficient system to deploy a network of autonomous drones to timely complete the designated monitoring task at a reasonable cost in such dynamic conditions. One important challenge in such systems is the task coordination between the deployed UAVs.
 
Several coalition formation approaches have been developed for task allocation 
in UAV networks \cite{Afghah_ACC18,Mousavi_INFOCOM18,Li_potentialgame,MOUSAVI_adhoc,Ruan_game_coalition}. A new approach is developing game-theoretic coalition formation approaches in a distributed manner, where the solution is obtained by considering the individual benefits of the agents as well as the social benefit of the entire network \cite{Afghah_ACC18,Li_potentialgame,Ruan_game_coalition}. In \cite{Mousavi_INFOCOM18}, a quantum-inspired genetic algorithm is proposed for leader-follower coalition formation that finds the optimum coalitions of UAVs in a large-scale system by accounting for reliability and probability of UAVs' failure. However, these methods require a considerable time to find the optimal set of coalitions, which can limit their applications in real world applications. Another common disadvantage of these coalition formation methods is the computational complexity of forming stable coalitions. In general, conventional coalition formation methods such as dynamic programming methods \cite{Rahwan,Cerquides}, graph theoretic approaches \cite{Sless,Bistaffa}, merge-and-split \cite{Afghah_ACC18} are not well suited to provide urgent responses in disaster monitoring applications in dynamic networks. 
In this paper, we propose a low-complexity fully distributed solution for this problem with close-to-optimal performance.


\section{Leader-follower Coalition Formation for UAV-based Wildfire Monitoring}
\begin{figure}[htb]
\centering
    \includegraphics[width=0.45\textwidth]{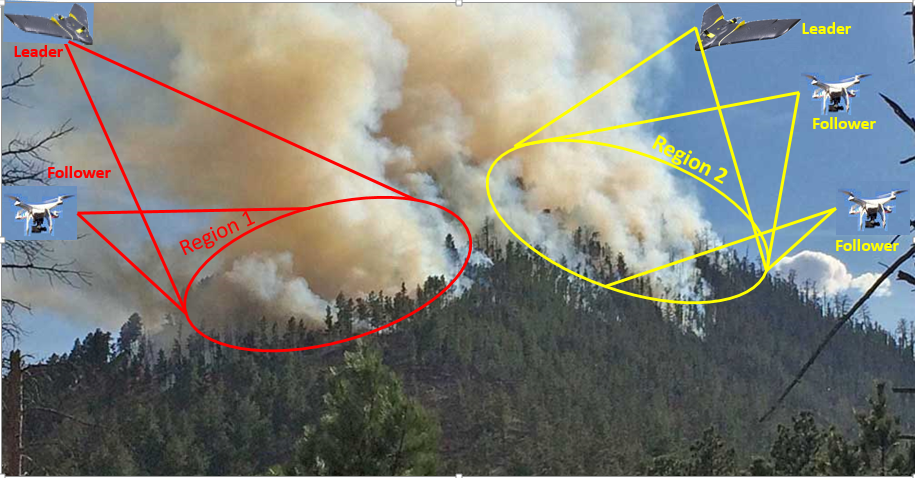}
\vspace{-0.3cm}
    \caption{An example of the proposed leader-follower UAV coalition formation in monitoring of two forest fire zones.}
    \label{fig:coalition}
\end{figure}
In this work, we propose a leader-follower coalition formation approach to monitor an active fire, specially in remote regions where deploying remotely controlled drones is challenging (see Figure~\ref{fig:coalition}). The main objective of the proposed solution is forming optimal UAV coalitions in a distributed manner to enable a full coverage of the fire field in a timely manner, when the observer UAVs have limited communication range and hence cannot directly send their collected information to the fire-management ground station. 

We consider a set of heterogeneous UAVs with different sensing/imaging capabilities, payload size, battery type, and flight time. 
We utilize two types of drones including \textit{fixed-wing} UAVs that can fly at higher attitudes and rapidly survey a wide area and the \textit{rotary} UAVs that hover at low altitudes to collect high resolution data. The fixed-wing UAVs with better flight capabilities and higher computation capabilities serve as \textit{coalition leaders} and are used for initial recognition of fires to relay the accumulated information within a coalition to the ground station. The rotary UAVs serve as \textit{coalition members} (or \textit{followers}) and perform the actual sensing and video recording tasks, hence called \textit{observer UAVs} in this work.

Depending on the incidence rate and severity of wildfires in a region, we can consider scenarios with different requirements in terms of the number and the type of UAVs. The UAVs can be situated in stand-by mode in lookout tower stations or on weather-resistant charging pads located across the forest. 
In the proposed event-triggered fire monitoring system, we assume that the UAVs' mission is initiated based on primary information about a fire provided by a human observer or a ground sensor grid. Then, the high-speed fixed-wing UAVs 
fly towards the impacted area to confirm the existence of fire and provide an initial fire map. 
After the preliminary evaluation of the fire region, the fixed-wing UAVs provide an approximate fire profile including an estimated thermal mapping of the fire, spread rate, flame length, fire intensity, and the estimated number and type of UAVs required to provide a full imaging/video coverage of the impacted region. Then, the following coalition formation process is initiated.  

\begin{figure}[htb]
\centerline{\includegraphics[width=0.8\linewidth]{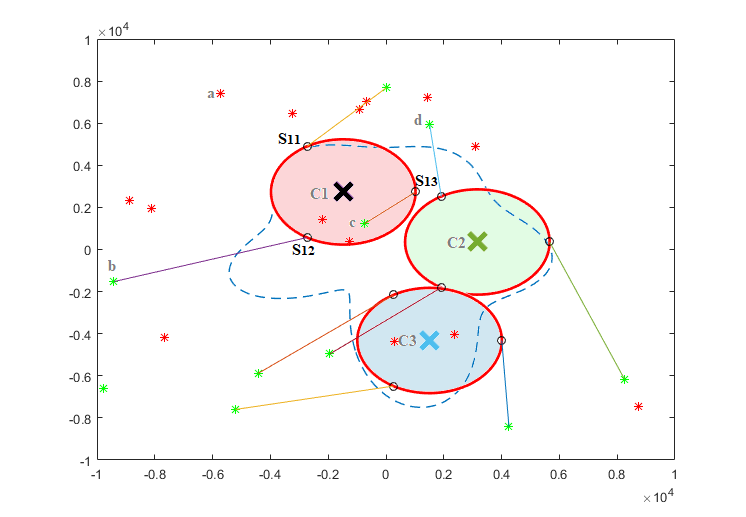}}
\vspace{-0.4cm}
\caption{An illustrative example of optimal coalition formation for maximal coverage. The leader UAV employs the member drones based on their flight time to the designated locations as well as their remaining batteries. The drones hover in a circular path to collect imagery information.}
\label{fig:coal1}
\end{figure}

Fig. \ref{fig:coal1} illustrates the operation of the proposed algorithm. The fire-zone area is shown by a dashed line. In the first stage, the leader UAVs take greedy displacement actions to relocate themselves to achieve two objectives of: i) creating at least 80 percent intersection among the respective coalition zones and the fire-zone, and ii) maintaining maximal separation between the coalition centers. These two conditions facilitate monitoring the active fire with more emphasis on the fire front line. Fig. \ref{fig:coal1} shows an example of the proposed leader-follower coalition formation system with three leaders that form three coalitions ($C=3$) denoted by $\mathcal{C}_1$, $\mathcal{C}_2$, and $\mathcal{C}_3$.

Once this phase is completed, the leader UAVs initiate the coalition formation process to recruit appropriate follower (observer) UAVs. Here, we use a sectorized approach, in which the coverage region of each leader UAV is divided to a predefined number of sectors, denoted by $S$. Each sector of a coverage region is covered by one observer drone (equipped with a high-resolution camera). The number of sectors is an application-dependent parameter and depends on the coalition coverage area, the communication range of observer UAVs, the UAVs' point of view, and the desired resolution and frequency of the captured images and videos. Here, we arbitrarily use $S=3$. The observer drones are maximally separated, therefore each sector represents an arc with a central angle of $2\pi/S$. The drones hover with a constant speed  in circular clockwise paths to collect video while avoiding potential collisions. 

The coalition formation is performed by a bid-response negotiation between the leaders and followers in a close proximity of one another. We should note that each coalition is formed independently by a leader, hence no communication is required among the leaders that can be located far apart from each other in different fire regions. Therefore, this coalition formation mechanism does not impose any signaling load to the leaders to coordinate their coalition formation.   

More formally, we consider a set of $N$ available follower UAVs, denoted by $\mathcal{U} = \{U_1,U_2,\dots,U_{N}\}$, where each follower UAV $U_i$ is described with an identification vector, $I_i$. This vector includes the information related to the UAV's available sensing and actuation features based on onboard sensor and actuator types, its consumable resources (e.g. remaining battery) and flying capabilities (maximum and minimum speed, allowable flight altitude). More specifically, we define $I_i = [ p_i^{1},p_i^{2}, \ldots, p_i^{N_p}, r_i^{1},r_i^{2}, \ldots, r_i^{{N_r}}]$, where $\textbf{p}_i= [p_i^{1},p_i^{2}, \ldots, p_i^{N_p}]$ with $p_i^j \geq 0$ represents the characteristic vector of user $i$ (e.g., maximum flight altitude, collision avoidance feature, and maximum tolerable temperature), and $\textbf{r}_i=[r_i^{1},r_i^{2}, \ldots, r_i^{{N_r}}]$ with $r_i^j \geq 0$ represents the available resources at UAV $U_i$. Notations $N_p$ and $N_r$ denote the maximum number of characteristics and resources, required to describe UAV $i$, respectively. The resources can be consumable such as remaining battery or non-consumable such as onboard sensor types. 
If $U_i$ does not have any of resource $j$, then $r_i^j = 0$ and if resource $j$ is non-consumable, then $r_i^j = \infty$. 

Each leader UAV, denoted by $L_k$, identifies (or assigned with) a set of spatially distributed tasks in its coverage region, presented by $\mathcal{T}_k=\{T_k^{i_1},T_k^{i_2},\ldots,T_k^{i_k}\}$, where $i_k$ is the number of identified tasks in that region. Each task $T_k^i$ requires a set of resources $R_k^i$ as well as properties $P_k^i$. Therefore, a coalition $C_k$ has a requirement vector $[P_k,R_k]$ with $P_k=\max \{P_k^1, P_k^2, \dots, P_k^{i_k}\}$ (with element-wise maximization) and $R_k=\sum_{j=1}^{i_k} R_k^j$.  If the leader UAV does not have sufficient resources to perform all these tasks in the required time frame, or if it cannot provide a full sensing/imaging coverage of the entire region, it calls for a coalition formation. Each coalition member (i.e. follower UAV) should have the required properties $P_k$ and the coalition members collectively should provide the required resources. More formally, if the coalition $C_k$ is formed by $\{U_{k_1},U_{k_2},\dots,U_{k_S}\}$ then, we should have:
\begin{align}
\nonumber
\mathbf{p}_j \geq P_k, &\text{ for }j=k_1,k_2,\dots,k_S,\\
&\text{subject to:} \sum_{j=k_1}^{k_S} \mathbf{r}_j \geq R_k.\vspace{-0.2cm}
\end{align}
In the proposed coalition formation model, first the leader of each region broadcasts a \emph{proposal} to the potential follower UAVs in its close proximity to form a coalition. This proposal includes information regarding the types of sensors or cameras that are needed, the approximate duration of the mission, as well as the coordinates of the fire region that it covers. Then, the UAVs which possess the required properties that the leader asked for respond to the request by reporting their properties (e.g., their flight capabilities), available resources (e.g. remained battery), and their current position. During the \emph{formation process}, the coalition leader evaluates all the responses by assessing the resources offered by the volunteer UAVs,  their remained battery, and their current location. 
From the leader' perspective, the recruited followers need to have the required properties to perform the detected tasks in the region, reach the fire region with minimal latency, and also collectively carry all the required resources for the encountered tasks in the region. Moreover, since in agile remote sensing applications such as disaster monitoring and forest fire monitoring, the number of available UAVs, and in particular the number of costly and high-capability UAVs is limited,  the leader UAVs need to account for a balanced distribution of the UAVs and their resources in the entire region. Meaning that coalition formation by the leader should be based on assuring that all the required resources for a region are provided by the followers while the resources do not considerably surpass the required resources for that region. Therefore, the objective of leaders in forming coalitions is to select the coalition members in such a way to: i) guarantee providing the required resources and capabilities to complete the tasks while not exceeding those, ii) guarantee the timely completion of the tasks, iii) choose the UAVs that are in closer distances to the region of interest, iv) select the ones with longer lifetime. In a simple scenario, this member selection for leader $k$ can be modeled as follows:\vspace{-0.2cm}
\begin{align} \label{eq:v}
\nonumber
& \text{Select the best follower UAVs to maximize:}\\
& ~~~~~~~~~~~~~v(\mathcal{C}_k) =  \sum\limits_{l = 1}^{N_r}\gamma (\sum\limits_{j=k_1}^{k_S}\mathbf{r}_j^l)/R_k^l), 
\end{align}
\vspace{-0.5cm}
\begin{align}
\nonumber
&\text{subject to:       }  ~~~\mathbf{p}_j \geq P_k, \text{ for }j=k_1,k_2,\dots,k_S,\\
\nonumber
&~~~~~~~~~~~~~~~d_{jk} \leq D.
\vspace{-5pt}
\end{align}
where $v(\mathcal{C}_k)$ denotes the value of coalition $\mathcal{C}_k$. The second constraint is to ensure that the distance of $U_j$ to the center of coalition $\mathcal{C}_k$, denoted by $d_{jk}$ is below a predefined threshold $D$ to ensure that the coalition formation time is not delayed longer than a desired time.  
The function $\gamma$ is designed to ensure that all the required resources for this region are provided by coalition members while over-spending is penalized. For example, the function can be defined as 
\begin{align}  \label{eq:penalty}
\gamma(x) = 
\begin{cases}
-L,     & \text{ if } x \leq 1,  \\
-x,    & \text{ if } x > 1 ,
\end{cases}
\end{align}
where $L \rightarrow \infty$ is a large number to harshly penalize coalitions with insufficient resources.

When the leader UAVs select an optimal set of their desired follower UAVs, they send a request to these UAVs to join the coalition. This request is called a \textit{bid}. Each follower UAV can receive multiple bids from different leaders. The potential followers will consider several factors to select the most appropriate offer from leader such as priority of regions, and its distance to the operation regions of the leaders. Then, the follower UAVs respond to the leaders' bids with a 'yes' or 'no' answer. If the leaders receive a 'yes' response from all the followers they originally selected, the coalition formation is completed. Otherwise, they will recruit the ones who sent a positive response and perform the member selection optimization process again to select additional required followers.  Once a coalition is formed, the leader assigns the selected members into $S$ sectors, such that the sum of the UAVs' flight time to their designated locations is minimized among all possible $S!$ permutations. Then, the leader instructs the UAVs to fly to the corresponding sectors and the task performance is initiated. During the lifetime of a mission, if a UAV collapses for any reasons (e.g., battery exhaustion or communication system failure), the coalition will recruit another follower to fill its position. If the leader UAV of a region becomes out of order, another UAV in the coalition will replace it (e.g., the follower with longest battery life).

\section{Simulation Results}
In this section, we present simulation results to evaluate the performance of the proposed algorithm in accommodating maximal coverage through forming coalitions of drones. In the simulations we set parameters as N=$20$ drones, $C=3$ coalitions and $S=3$ sectors. We also use a randomly generated fire zone in a rectangular service area with side $L=10 km$. The drones' affordable flight time based on their batteries are uniformly distributed between $10~min$ and $20~min$, while the mission time is considered as $15~min$, therefore the drones  that their remaining battery after flying to the designated destinations is sufficient to perform the task are included in the coalition formation (this condition is imposed by $d_{jk}\leq D$ in (\ref{eq:v})). 


\begin{figure}[hbt!]
    \centering
    \begin{subfigure}[t]{0.49\textwidth}
        \centering
  	    \includegraphics[width=1\textwidth]{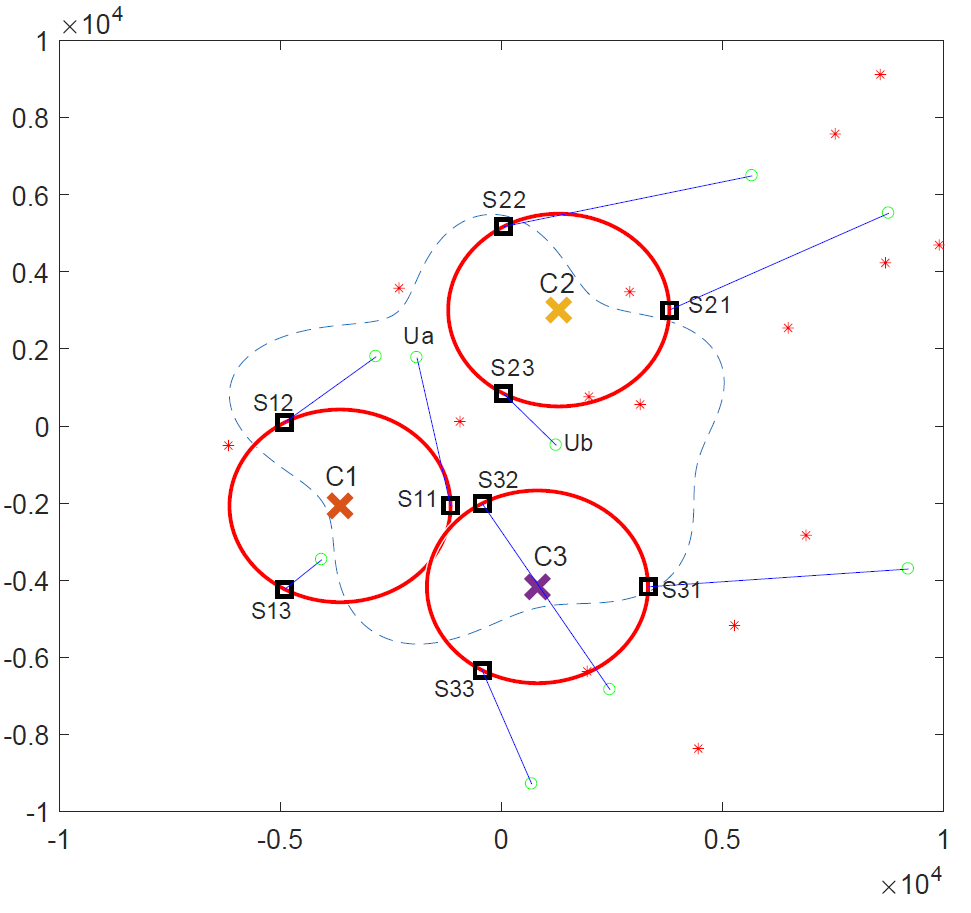}
    \end{subfigure}
    \\
    ~ 
    \begin{subfigure}[t]{0.49\textwidth}
        \centering
  	    \includegraphics[width= 1\textwidth]{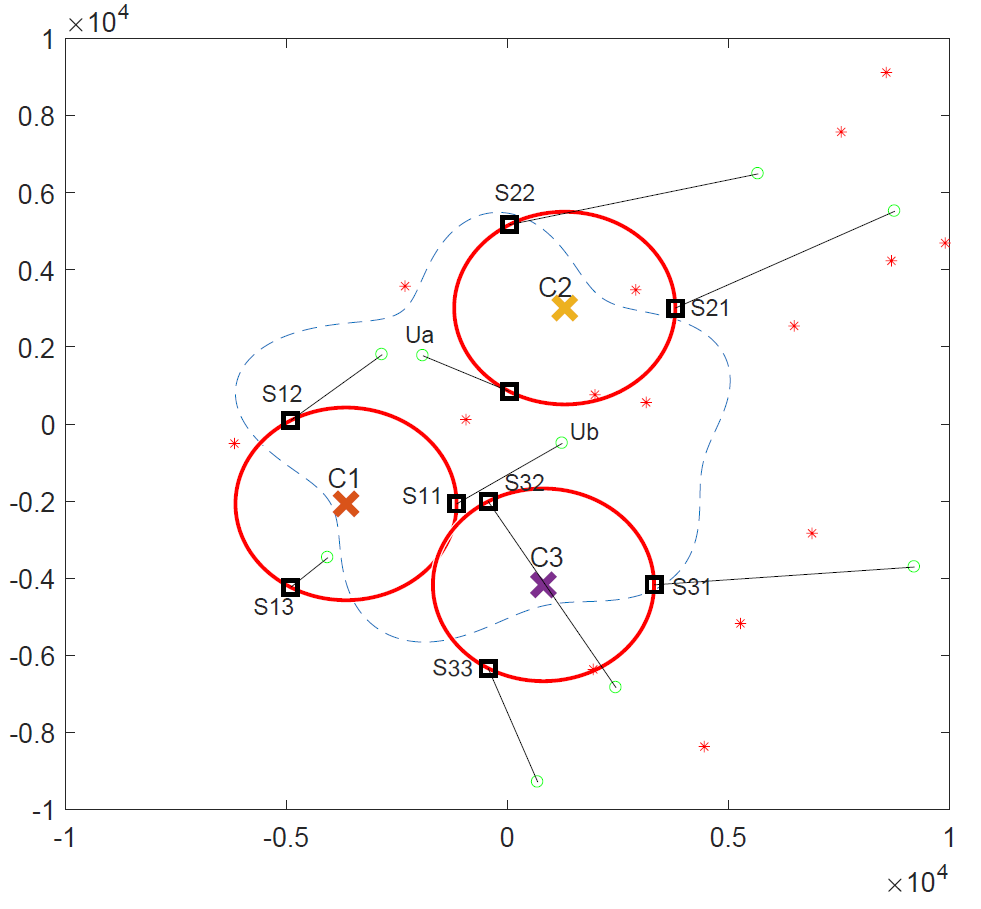}
    \end{subfigure}
    \caption{Optimal coalition formation for maximal coverage. \textbf{top:} the coalition formation using central method;  \textbf{bottom:} the coalition formation using the proposed distributed leader-follower method; The drones whose property vectors satisfy condition $\mathbf{p}_j\geq P_k$ in (1) are shown with green circles, and the rest of drones are marked with red stars.}
    \label{fig:comp1}
\end{figure}

In order to illustrate the operation of the proposed algorithm, we compared the resulting coalitions with a centralized optimized method. 
In Fig. \ref{fig:comp1}, the top figure is the formed coalitions using a centralized method, where the leaders are allowed to convene and develop optimal coalitions after sharing their information. In other words, the leaders examine the performance of all possible assignments of $N=24$ UAVs and $C = 3\times 3=9$ positions, namely the $N! / (C \times S)!$. Then, a mapping $\Psi: N \rightarrow C\times S$ that maximizes the summation of coalition values $\sum_{c=1}^c v(\mathcal{C}_c)$ is selected, where $v(\mathcal{C}_c)$ is defined in (\ref{eq:v}). The central optimization requires inter-coalition communications that may not be practical in real situations. 
The bottom figure in Fig. \ref{fig:comp1} illustrates the coalitions formed by the proposed fully distributed algorithm, where the leaders form their coalitions independently after negotiating to nearby observer UAVs. This figure shows that the final coalitions are very close to the solution of centralized algorithm. The only exception is the position of drones $U_a$ and $U_b$ that their positioned sectors $S_{11}$ and $S_{23}$ are swapped between the two methods. An intuitive justification is that in the proposed distributed method, the leader of coalition $\mathcal{C}_1$ selects $U_b$ for position $S_{11}$ first, and then the leader of coalition $\mathcal{C}_2$ has no option better than $U_a$ for position $S_{23}$, where as the central one selects a better assignment of $U_a \rightarrow S_{11}$ and $U_b\rightarrow S_{23}$.

\begin{figure}[htb]
\centerline{\includegraphics[width=0.8\linewidth]{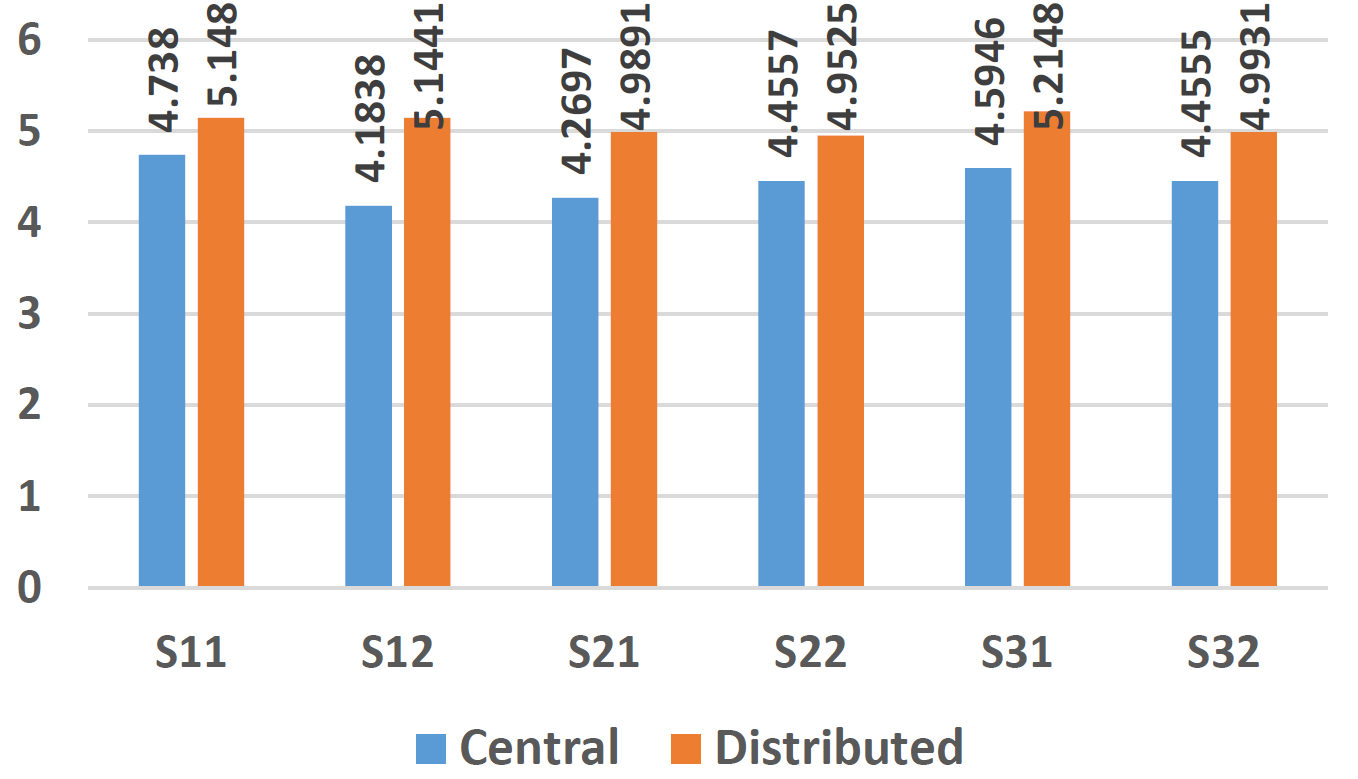}}
\vspace{-0.3cm}
\caption{The average distance between coalition members' initial positions to their target sectors for both centralized and distributed methods. The average is taken over 100 runs of the algorithm with random initialization.}
\label{fig:comp2}
\end{figure}

 To further investigate the performance of the system, we execute 100 scenarios with randomly generated fields and parameters $L=10^4, N=15, C=3$, and $S=2$. We use remaining battery as the only resource and hence the maximization in (\ref{eq:v}) is equal to minimizing the distance between the selected coalition members with their designated locations. The results in Fig. \ref{fig:comp2} illustrate that the central algorithm performs consistently better than the distributed algorithm across all sector positions as expected. But, the difference between the average distances for the two methods is as low as $15\%$.

\section{conclusion}
We developed a practical model for disaster monitoring in remote regions using multiple autonomous UAVs. It is assumed that several UAVs with different flight and sensing capabilities are situated in lookout tower stations or charging in an stand-by mode. When a fire incident is reported by smoke sensors or human observers, the autonomous UAVs initiate a monitoring mission without human intervention. This monitoring mission is performed using a leader-follower coalition formation approach, in which the UAVs with longer flight time, higher communication ranges, and more computation powers serve as the leaders to form a set of coalitions that can monitor the fire zone with an expected quality and for the required duration. The simulation results show among hundreds of scenarios confirm that the proposed fully distributed algorithm performs close to a central method, while enabling a full autonomy and eliminating the need for inter-leader communication and coordination. As such, the system is scalable to scenarios with vast coverage areas, including forest fires.

\section{ACKNOWLEDGMENT OF SUPPORT AND DISCLAIMER}
The authors acknowledge the U.S. Government’s support in the publication of this paper. This material is in part based upon work funded by AFRL. Any opinions, findings and conclusions or recommendations expressed in this material are those of the author(s) and do not necessarily reflect the views of the US government or AFRL.

\bibliography{main}
\bibliographystyle{IEEEtran}

\end{document}